\documentclass[letterpaper,twocolumn,prl,aps,superscriptaddress,floatfix]{revtex4}

\usepackage{xcolor}
\usepackage{graphicx}
\usepackage{hyperref}
\usepackage{upgreek}
\hypersetup{hidelinks,colorlinks=true,allcolors=black,pdfstartview=Fit,breaklinks=true}

\begin{document}

\title{Experimental observation of spin defects in van der Waals material GeS$_2$}

\affiliation{CAS Key Laboratory of Quantum Information, University of Science and Technology of China, Hefei, Anhui 230026, China}
\affiliation{Anhui Province Key Laboratory of Quantum Network, Hefei, 230026, China}
\affiliation{HUN-REN Wigner Research Centre for Physics, P.O.\ Box 49, H-1525 Budapest, Hungary}
\affiliation{Beijing Computational Science Research Center, Beijing 100193, China}
\affiliation{CAS Center For Excellence in Quantum Information and Quantum Physics, University of Science and Technology of China, Hefei, 230026, China}
\affiliation{Hefei National Laboratory, University of Science and Technology of China, Hefei 230088, China.}
\affiliation{Department of Atomic Physics, Institute of Physics, Budapest University of Technology and Economics,  M\H{u}egyetem rakpart 3., H-1111 Budapest, Hungary}
\affiliation{MTA–WFK Lend\"{u}let "Momentum" Semiconductor Nanostructures Research Group, P.O.\ Box 49, H-1525 Budapest, Hungary}

\author{Wei Liu}
\thanks{These authors contributed equally to this work.}
\affiliation{CAS Key Laboratory of Quantum Information, University of Science and Technology of China, Hefei, Anhui 230026, China}
\affiliation{Anhui Province Key Laboratory of Quantum Network, Hefei, 230026, China}
\affiliation{CAS Center For Excellence in Quantum Information and Quantum Physics, University of Science and Technology of China, Hefei, 230026, China}

\author{Song Li}
\thanks{These authors contributed equally to this work.}
\affiliation{HUN-REN Wigner Research Centre for Physics, P.O.\ Box 49, H-1525 Budapest, Hungary}
\affiliation{Beijing Computational Science Research Center, Beijing 100193, China}

\author{Nai-Jie Guo}
\affiliation{CAS Key Laboratory of Quantum Information, University of Science and Technology of China, Hefei, Anhui 230026, China}
\affiliation{Anhui Province Key Laboratory of Quantum Network, Hefei, 230026, China}
\affiliation{CAS Center For Excellence in Quantum Information and Quantum Physics, University of Science and Technology of China, Hefei, 230026, China}
\affiliation{Hefei National Laboratory, University of Science and Technology of China, Hefei 230088, China.}

\author{Xiao-Dong Zeng}
\affiliation{CAS Key Laboratory of Quantum Information, University of Science and Technology of China, Hefei, Anhui 230026, China}
\affiliation{Anhui Province Key Laboratory of Quantum Network, Hefei, 230026, China}
\affiliation{CAS Center For Excellence in Quantum Information and Quantum Physics, University of Science and Technology of China, Hefei, 230026, China}

\author{Lin-Ke Xie}
\affiliation{CAS Key Laboratory of Quantum Information, University of Science and Technology of China, Hefei, Anhui 230026, China}
\affiliation{Anhui Province Key Laboratory of Quantum Network, Hefei, 230026, China}
\affiliation{CAS Center For Excellence in Quantum Information and Quantum Physics, University of Science and Technology of China, Hefei, 230026, China}

\author{Jun-You Liu}
\affiliation{CAS Key Laboratory of Quantum Information, University of Science and Technology of China, Hefei, Anhui 230026, China}
\affiliation{Anhui Province Key Laboratory of Quantum Network, Hefei, 230026, China}
\affiliation{CAS Center For Excellence in Quantum Information and Quantum Physics, University of Science and Technology of China, Hefei, 230026, China}
\affiliation{Hefei National Laboratory, University of Science and Technology of China, Hefei 230088, China.}

\author{Yu-Hang Ma}
\affiliation{CAS Key Laboratory of Quantum Information, University of Science and Technology of China, Hefei, Anhui 230026, China}
\affiliation{Anhui Province Key Laboratory of Quantum Network, Hefei, 230026, China}
\affiliation{CAS Center For Excellence in Quantum Information and Quantum Physics, University of Science and Technology of China, Hefei, 230026, China}

\author{Ya-Qi Wu}
\affiliation{CAS Key Laboratory of Quantum Information, University of Science and Technology of China, Hefei, Anhui 230026, China}
\affiliation{Anhui Province Key Laboratory of Quantum Network, Hefei, 230026, China}
\affiliation{CAS Center For Excellence in Quantum Information and Quantum Physics, University of Science and Technology of China, Hefei, 230026, China}

\author{Yi-Tao Wang}
\affiliation{CAS Key Laboratory of Quantum Information, University of Science and Technology of China, Hefei, Anhui 230026, China}
\affiliation{Anhui Province Key Laboratory of Quantum Network, Hefei, 230026, China}
\affiliation{CAS Center For Excellence in Quantum Information and Quantum Physics, University of Science and Technology of China, Hefei, 230026, China}

\author{Zhao-An Wang}
\affiliation{CAS Key Laboratory of Quantum Information, University of Science and Technology of China, Hefei, Anhui 230026, China}
\affiliation{Anhui Province Key Laboratory of Quantum Network, Hefei, 230026, China}
\affiliation{CAS Center For Excellence in Quantum Information and Quantum Physics, University of Science and Technology of China, Hefei, 230026, China}

\author{Jia-Ming Ren}
\affiliation{CAS Key Laboratory of Quantum Information, University of Science and Technology of China, Hefei, Anhui 230026, China}
\affiliation{Anhui Province Key Laboratory of Quantum Network, Hefei, 230026, China}
\affiliation{CAS Center For Excellence in Quantum Information and Quantum Physics, University of Science and Technology of China, Hefei, 230026, China}

\author{Chun Ao}
\affiliation{CAS Key Laboratory of Quantum Information, University of Science and Technology of China, Hefei, Anhui 230026, China}
\affiliation{Anhui Province Key Laboratory of Quantum Network, Hefei, 230026, China}
\affiliation{CAS Center For Excellence in Quantum Information and Quantum Physics, University of Science and Technology of China, Hefei, 230026, China}

\author{Jin-Shi Xu}
\affiliation{CAS Key Laboratory of Quantum Information, University of Science and Technology of China, Hefei, Anhui 230026, China}
\affiliation{Anhui Province Key Laboratory of Quantum Network, Hefei, 230026, China}
\affiliation{CAS Center For Excellence in Quantum Information and Quantum Physics, University of Science and Technology of China, Hefei, 230026, China}
\affiliation{Hefei National Laboratory, University of Science and Technology of China, Hefei 230088, China.}

\author{Jian-Shun Tang}
\email{tjs@ustc.edu.cn}
\affiliation{CAS Key Laboratory of Quantum Information, University of Science and Technology of China, Hefei, Anhui 230026, China}
\affiliation{Anhui Province Key Laboratory of Quantum Network, Hefei, 230026, China}
\affiliation{CAS Center For Excellence in Quantum Information and Quantum Physics, University of Science and Technology of China, Hefei, 230026, China}
\affiliation{Hefei National Laboratory, University of Science and Technology of China, Hefei 230088, China.}

\author{Adam Gali}
\email{gali.adam@wigner.hun-ren.hu}
\affiliation{HUN-REN Wigner Research Centre for Physics, P.O.\ Box 49, H-1525 Budapest, Hungary}
\affiliation{Department of Atomic Physics, Institute of Physics, Budapest University of Technology and Economics,  M\H{u}egyetem rakpart 3., H-1111 Budapest, Hungary}
\affiliation{MTA–WFK Lend\"{u}let "Momentum" Semiconductor Nanostructures Research Group, P.O.\ Box 49, H-1525 Budapest, Hungary}

\author{Chuan-Feng Li}
\email{cfli@ustc.edu.cn}
\affiliation{CAS Key Laboratory of Quantum Information, University of Science and Technology of China, Hefei, Anhui 230026, China}
\affiliation{Anhui Province Key Laboratory of Quantum Network, Hefei, 230026, China}
\affiliation{CAS Center For Excellence in Quantum Information and Quantum Physics, University of Science and Technology of China, Hefei, 230026, China}
\affiliation{Hefei National Laboratory, University of Science and Technology of China, Hefei 230088, China.}

\author{Guang-Can Guo}
\affiliation{CAS Key Laboratory of Quantum Information, University of Science and Technology of China, Hefei, Anhui 230026, China}
\affiliation{Anhui Province Key Laboratory of Quantum Network, Hefei, 230026, China}
\affiliation{CAS Center For Excellence in Quantum Information and Quantum Physics, University of Science and Technology of China, Hefei, 230026, China}
\affiliation{Hefei National Laboratory, University of Science and Technology of China, Hefei 230088, China.}

\begin{abstract}
    Spin defects in atomically thin two-dimensional (2D) materials such as hexagonal boron nitride (hBN)  attract significant attention for their potential quantum applications. The layered host materials not only facilitate seamless integration with optoelectronic devices but also enable the formation of heterostructures with on-demand functionality. Furthermore, their atomic thickness renders them particularly suitable for sensing applications. 
    However, the short coherence times of the spin defects in hBN limit them in quantum applications that require extended coherence time. 
    One primary reason is that both boron and nitrogen atoms have non-zero nuclear spins. 
    Here, we present another 2D material germanium disulfide ($\upbeta$-GeS$_2$) characterized by a wide bandgap and potential nuclear-spin-free lattice. This makes it as a promising host material for spin defects that possess long-coherence time. 
    Our findings reveal the presence of more than two distinct types of spin defects in single-crystal $\upbeta$-GeS$_2$. 
    Coherent control of one type defect has been successfully demonstrated at both 5 K and room temperature, and the coherence time $T_2$ can achieve tens of microseconds, 100-folds of that of negatively charged boron vacancy (V$_{\text{B}}^-$) in hBN, satisfying the minimal threshold required for metropolitan quantum networks--one of the important applications of spins. We tentatively assign the observed optical signals come from substitution defects. Together with previous theoretical prediction, 
    we believe the coherence time can be further improved with optimized lattice quality, indicating $\upbeta$-GeS$_2$ as a promising host material for long-coherence-time spins.
\end{abstract}

\maketitle

Optically active spins based on defects in solids are promising platforms for quantum technology, owning to their efficient spin-photon interface and scalability \cite{Bernien2013Heralded,Humphreys2018Deterministic,Pompili2021Realization,Bhaskar2020Experimental,Stas2022Robust,Knaut2024Entanglement,Togan2010Quantum,Fang2023Experimental,Cai2013A,Zu2014Experimental,Rong2015Experimental,Wu2019Observation,Wu2024Third,Shi2015Single,Wang2023Magnetic}. 
The nitrogen-vacancy (NV$^-$) and silicon-vacancy (SiV$^-$) centers in diamond, divacancy ($\rm V_{Si}–V_C$) and silicon vacancy ($\rm V_{Si}$) in silicon carbide (SiC) are the most prominent color centers, which have been used to demonstrate quantum network \cite{Bernien2013Heralded,Humphreys2018Deterministic,Pompili2021Realization,Bhaskar2020Experimental,Stas2022Robust,Knaut2024Entanglement}, quantum simulation and computation \cite{Cai2013A,Zu2014Experimental,Rong2015Experimental,Wu2019Observation,Wu2024Third}, quantum sensing and metrology \cite{Shi2015Single,Wang2023Magnetic}, etc. 
Recently, spin defects in two-dimensional (2D) van der Waals materials have been attracting enormous attention \cite{Liu2022Spin,Fabian2022Quantum,Sajid2022Spin}. The atomic-scale thickness and various growth methods make it easy for 2D materials to be integrated with other quantum architectures and bring up additional functionalities. 
Defects inside hence possess high photon-extraction efficiency and are inherently near surface, allowing them easier to be created, characterized and manipulated. 
Some optically addressable spin defects have been experimentally discovered in hexagonal boron nitride (hBN) \cite{Gottscholl2020Initialization,Mendelson2021Identifying,Chejanovsky2021Single,Stern2022Room,Guo2023Coherent,Yang2023Laser,Stern2024A,Scholten2024Multi}. 
Among them, the extensively studied negatively charged boron vacancy ($\rm V_B^-$) in hBN has achieved excellent performance in quantum sensing of magnetic field, temperature, nuclear spin, paramagnetic spins in liquids, and beyond \cite{Vaidya2023Quantum,Gottscholl2021Spin,Liu2021Temperature,Healey2023Quantum,Huang2022Wide,Kumar2022Magnetic,Zhou2023DC,Gao2022Nuclear,Robertson2023Detection,Gao2023Quantum,Durand2023Optically}.

The spin coherence time $T_2$ of spin defect, generally measured by Hahn-echo experiment, characterizes the retention time of the quantum information.  
Applications including quantum network and quantum sensing based on Ramsey interferometry have high requirements for $T_2$. 
For example, quantum network \cite{Bernien2013Heralded,Humphreys2018Deterministic,Pompili2021Realization,Bhaskar2020Experimental,Stas2022Robust,Knaut2024Entanglement,Togan2010Quantum,Fang2023Experimental,Kimble2008The,Wehner2018Quantum,Awschalom2021Development,Jiang2007Distributed,Gottesman2012Longer,Komar2014A,Khabiboulline2019Optical,Liu2024Creation,Liu2023Distributed}, which can enable a range of other applications such as secure communication and distributed quantum computation/sensing, require $T_2$ to be long enough to allow photons to reach distant destinations while preserving quantum information. 
For basic fiber-connected metropolitan quantum networks, the general communications between adjacent subway stations ($\sim$kilometers) require coherence time reaching 10 $\upmu$s. 
However, the coherence times of spin defects in hBN are fairly short (e.g., $T_2\sim$100 ns for $\rm V_B^-$) \cite{Haykal2022Decoherence,Liu2022Coherent}, even under coherence protection schemes they were still measured to be only several microsecond \cite{Ramsay2023Coherence,Gong2023Coherent,Rizzato2023Extending}.
Therefore, it is imperative to find spin defects with excellent spin coherence in 2D materials.

\begin{figure*}[htp]
\centering
\includegraphics[width=0.8\textwidth]{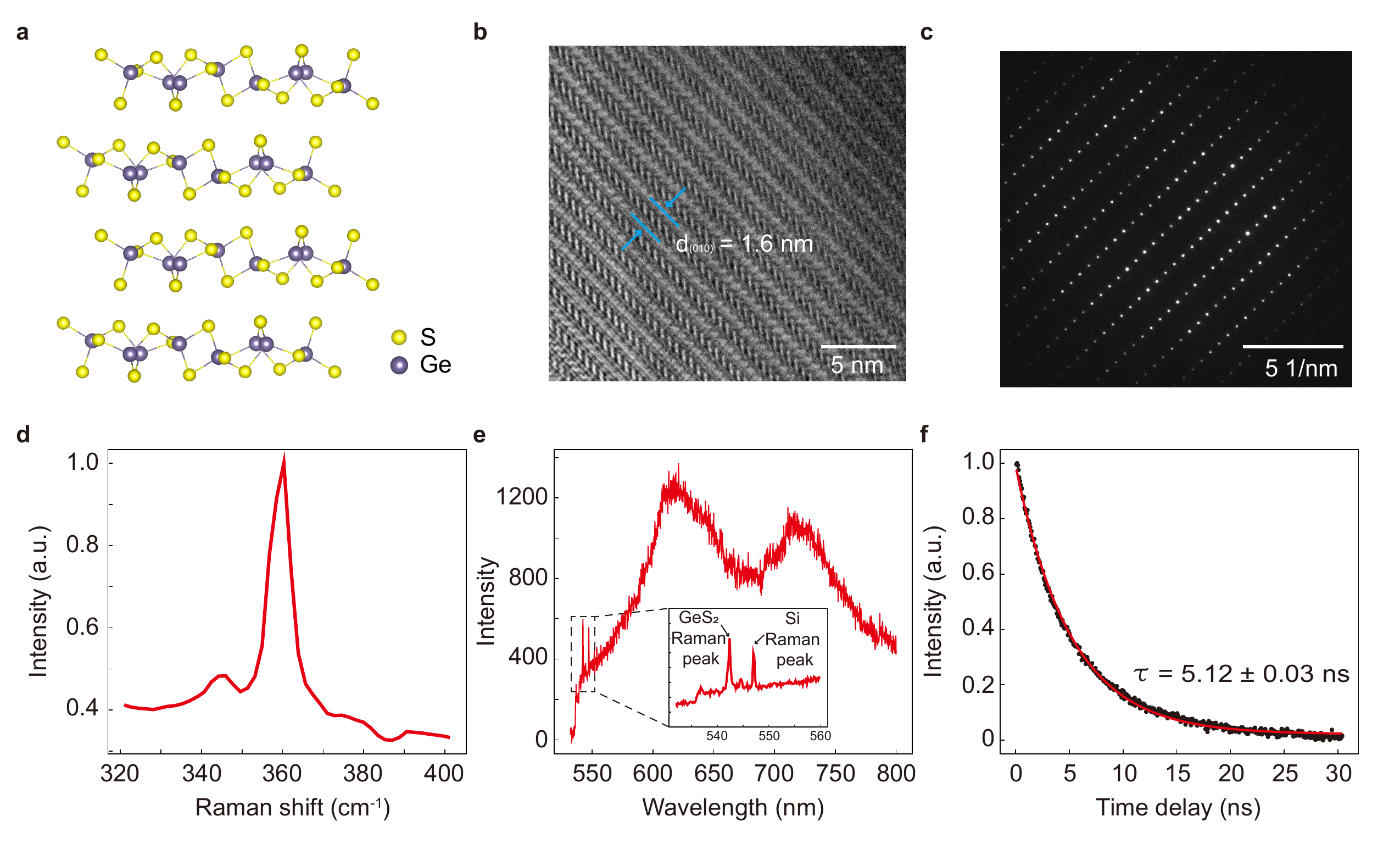}
\caption{\label{Figure 1} \textbf{Structural characterization of $\upbeta$-GeS$_2$ and optical characterization of the defects.} \textbf{a} Crystal structure of $\upbeta$-GeS$_2$ from side view. \textbf{b} HRTEM image of $\upbeta$-GeS$_2$ flake and \textbf{c} corresponding SEAD pattern. \textbf{d} Experimental Raman spectrum of a $\upbeta$-GeS$_2$ flakes. \textbf{e} Room temperature spectrum of color centers in $\upbeta$-GeS$_2$ flake. The Raman peaks of $\upbeta$-GeS$_2$ and Si are shown enlarged in the inset. \textbf{f} The fluorescence lifetimes of the defects at 5 K. The solid line is fitting result with the lifetime of 5.12$\pm$0.03 ns.}
\end{figure*}

There exists diverse sources that cause spin decoherence. In the case of defect spin, the coupling between the electron spin and the surrounding nuclear spin bath is the predominate reason. 
Since all the naturally existed boron ($^{10}$B and $^{11}$B) and nitrogen ($^{14}$N and $^{15}$N) have non-zero nuclear spins, ubiquitous and dense nuclear spin bath in the hBN lattice not only limits the upper bound of coherence time of electron spins \cite{Kanai2022Generalized}, but also makes the conventional isotopic purification ineffective \cite{Haykal2022Decoherence,Lee2022First,Clua2023Isotopic,Gong2024Isotope}.
Similarly, recently discovered spin defects in gallium nitride (GaN) \cite{Luo2024Room,Eng2024Room} also suffer from short coherence time ($\sim$100 ns) \cite{Luo2024Room}.
In contrast, the NV$^-$ center in diamond ($^{12}$C has a nuclear spin of $I$ = 0, 98.9\% natural abundance; $^{13}$C has a nuclear spin of $I$ = 1/2, 1.1\% natural abundance) has ultralong coherence time, and can further reach 1.8 ms at room temperature after isotope purification \cite{Balasubramanian2009Ultralong}.
Based on theoretical calculation \cite{Kanai2022Generalized}, it is now much easier to screen out potential materials to host defects with long spin coherence time, such as germanium disulfide (GeS$_2$), germanium diselenide (GeSe$_2$), Tellurium dioxide (TeO$_2$) and so on.

In this work, we studied the visible color centers in $\upbeta$-GeS$_2$ which is a wide bandgap ($\sim$3.7 eV) layered material \cite{Yang2019Polarization,Yang20222D}. 
Using optically detected magnetic resonance (ODMR) technologies and external magnetic field, we find more than two types of room-temperature optically addressable spin defects. 
Furthermore, the coherent manipulations of the spin defects at both 5 K and room temperature has been successfully demonstrated, and the coherence times $T_2$ are 10.03$\pm$1.52 $\upmu$s and 20.89$\pm$9.59 $\upmu$s, respectively, $\sim$100 folds of that of $\rm V_B^-$ in hBN, reaching the minimal threshold required for metropolitan quantum network. 

\begin{figure*}[htp]
\centering
\includegraphics[width=0.7\textwidth]{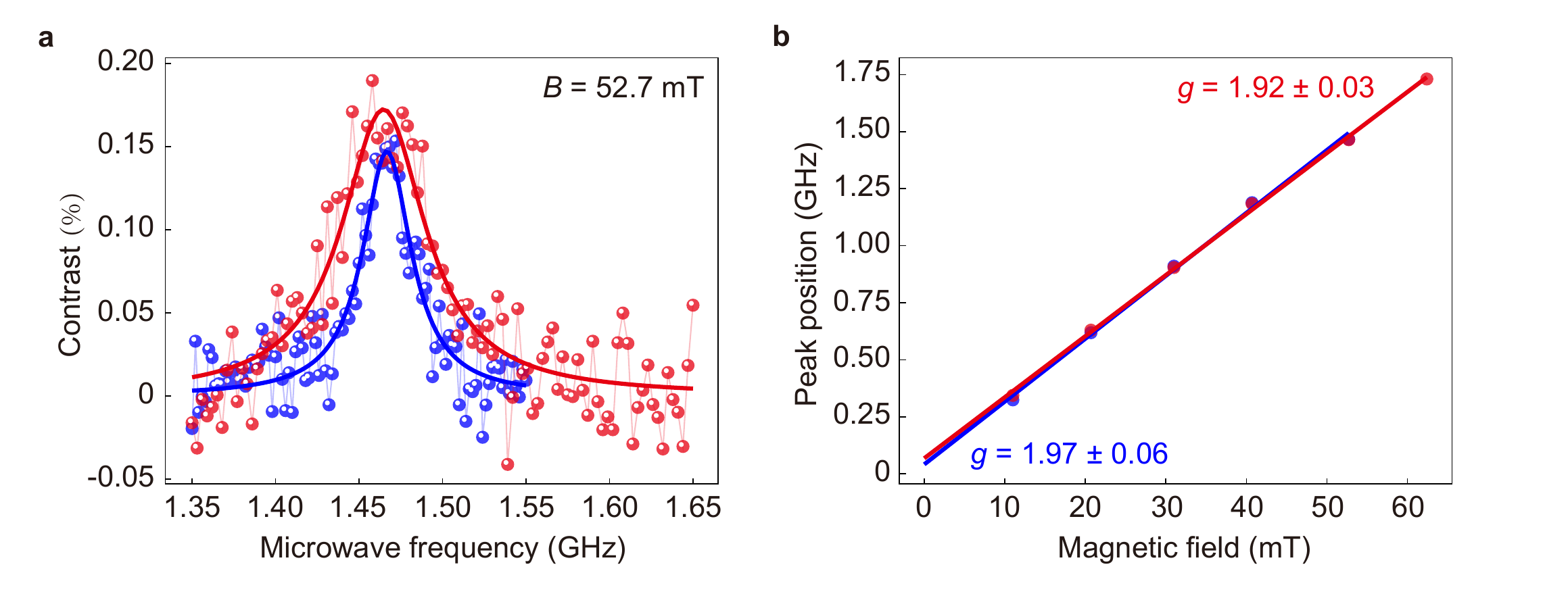}
\caption{\label{Figure 2} \textbf{Room-temperature optically detected electron spin resonance.}  \textbf{a} Two types of ODMR spectra of spin defects in $\upbeta$-GeS$_2$ at external magnetic field ($\textbf{B} || \textbf{c}$, $B = 52.7$ mT). The solid lines are Lorentzian fitting results, displaying that the FWHMs of ODMR spectra are 35.4$\pm$2.1 MHz (blue, corresponding Group I) and 69.0$\pm$4.0 MHz (red, corresponding Group II), respectively. \textbf{b} Dependence of ODMR resonance frequencies on the magnetic field ($\textbf{B} || \textbf{c}$). The dots are experimental results extracted from Fig. S4 in Supplementary Information, based on which we fit the $g$ factor to be 1.97$\pm$0.06 for Group I (blue), and 1.92$\pm$0.03 for Group II (red). All experiments were conducted under 1-mW laser excitation and 100-mW microwave power.}
\end{figure*}

\textbf{Structural and optical characterization.}
$\upbeta$-GeS$_2$ is a low-symmetry 2D material with monoclinic structure (space group $P2_1/c$, $a$ = 6.72, $b$ = 16.10, $c$ = 11.44 $\rm \AA$) \cite{Popovic1996High,Yang2019Polarization,Wang2020Sub,Yan2022Investigation}. 
$\upbeta$-GeS$_2$ has bandgap around 3.7 eV \cite{Yang2019Polarization,Yang20222D} and weak interlayer coupling \cite{Yan2022Investigation}. 
Due to the in-plane anisotropic crystal structure, $\upbeta$-GeS$_2$ shows strong in-plane anisotropy in electrical, optical, and mechanical properties \cite{Wang2020Sub}, and has been used for polarization-dependent photoresponse in the ultraviolet (UV) region \cite{Yang2019Polarization}. 
Additionally, most of the constituent atoms of $\upbeta$-GeS$_2$ are nuclear spinless ($^{73}$Ge with a spin $I$ = 9/2, 7.7\% natural abundance, other isotopes with spins $I$ = 0; $^{33}$S with a spin $I$ = 3/2, 0.8\% natural abundance, other isotopes with spins $I$ = 0). 
All the above mentioned properties make $\upbeta$-GeS$_2$ an ideal material to host spin defects with long coherence time, and theoretical calculations show it can reach 4.3 ms \cite{Kanai2022Generalized}.

The single-crystal $\upbeta$-GeS$_2$ samples studied here are mechanically exfoliated from commercial bulk crystal. 
Fig. 1\textbf{a} displays the atomic structure of $\upbeta$-GeS$_2$, which is distortedly connected by basic building blocks of GeS$_4$ tetrahedra. 
The high-resolution transmission electron microscopy (HRTEM) image (Fig. 1\textbf{b}) demonstrates an evident lattice period of 1.6 nm corresponding to the (010) plane of $\upbeta$-GeS$_2$, and the selected-area electron diffraction pattern (SEAD) is shown in Fig. 1\textbf{c}. 
We apply Raman spectrometer to measure the mechanically exfoliated $\upbeta$-GeS$_2$ flakes at room temperature, and there is an obvious Raman peak at 361.4 cm$^{-1}$ (Fig. 1\textbf{d}). 
The above material measurement results are consistent with previous reports \cite{Yang2019Polarization,Yan2022Investigation}, confirming the samples studied here are $\upbeta$-GeS$_2$.

The exfoliated $\upbeta$-GeS$_2$ flakes are prepared onto silicon (Si) wafer with 280-nm thermally-grown SiO$_2$ toplayer, and then anneal in air at 500 $^\circ$C for 1 hour to active and stabilize the color centers. 
Under 532-nm laser excitation, annealed $\upbeta$-GeS$_2$ flakes exhibit prevalent bright photoluminescence (PL). 
All photoluminescent samples have similar spectra, and a typical room-temperature spectrum is shown in Fig. 1\textbf{e}, which is measured by a home-built confocal scanning microscopic system combined with a spectrometer. 
There are two large packets centered at 615 nm and 720 nm, respectively, and no obvious zero-phonon line (ZPL) observed. 
 Two sharp peaks (542.4 nm and 547.1 nm) can be seen in the inset of Fig. 1\textbf{e}, and they correspond to the Raman peaks of $\upbeta$-GeS$_2$ and Si, respectively, indicating that the PL does come from the $\upbeta$-GeS$_2$ sample. 
In addition, we measure the low-temperature fluorescence properties of the $\upbeta$-GeS$_2$ samples by using a liquid-helium-free cryostat.
The cryogenic spectra of these color centers are essentially same as the room-temperature spectrum, except for the small peaks in the range of 580-590 nm (see Fig. S2 in Supplementary Information). 
We also perform time-resolved PL decay experiment using time-correlated single photon counting (TCSPC) system. 
A typical fluorescence lifetime of these color centers at 5 K is 5.12$\pm$0.03 ns, as shown in Fig. 1\textbf{f}. 
Further temperature-dependent fluorescence lifetime measurements (see Fig. S3\textbf{b} in Supplementary Information) show that temperature has no major impact on the excited state lifetime of these color centers.

\textbf{ODMR spectra and coherent control.} 
The $\upbeta$-GeS$_2$ samples are then transferred onto a gold-film microwave waveguide to study the spin properties at room temperature by ODMR technologies. 
Under the external magnetic field perpendicular to the $\upbeta$-GeS$_2$ surface ($\textbf{B} || \textbf{c}$) provided by an electromagnet, some sample areas exhibit ODMR spectra with a single positive peak at room temperature.
The measured ODMR spectra fall into two groups, and Fig. 2\textbf{a} displays the typical ODMR spectra of these two groups measured under 1-mW, 532-nm laser excitation, 100-mW microwave power and 52.7-mT magnetic field. 
The Lorentzian fitting results show that these two groups of ODMR spectra have full widths at half maximum (FWHMs) of 35.4$\pm$2.1 MHz (blue, corresponding to Group I) and 69.0$\pm$4.0 MHz (red, corresponding to Group II), respectively.
More ODMR spectra of Group I and II at different magnetic fields are measured (see Fig. S4 in Supplementary Information), and the relations of the ODMR peak position and the magnetic field strength are shown in Fig. 2\textbf{b}. 
The fitted $g$-factors are 1.97$\pm$0.06 for Group I (blue), and 1.92$\pm$0.03 for Group II (red), both closed to 2, confirming it is from electronic spins.
Additionally, microwave-power-dependent ODMR spectra (see Fig. S5 in Supplementary Information) show no obvious hyperfine structure even at low microwave power, so the defect should be nuclear-spin-free. 

\begin{figure*}[htp]
\centering
\includegraphics[width=0.9\textwidth]{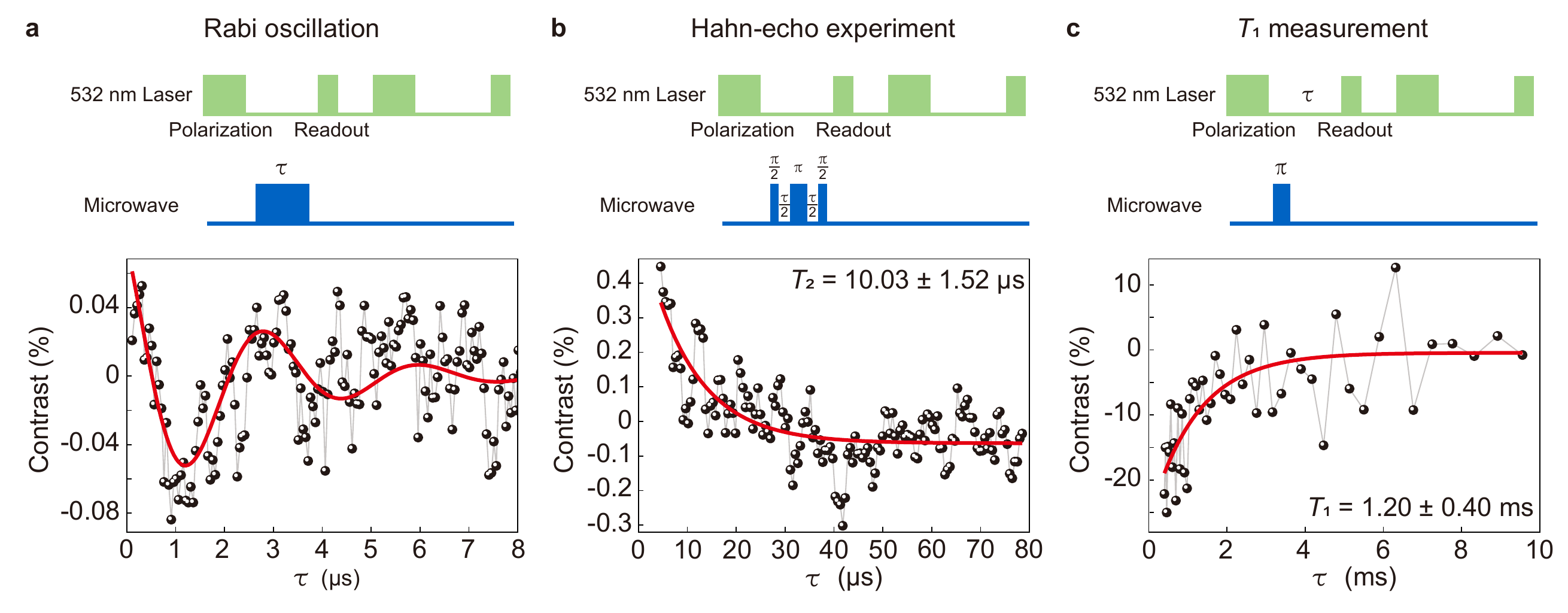}
\caption{\label{Figure 3} \textbf{Coherent manipulation of spin defects in $\upbeta$-GeS$_2$ (1-hour annealing) at 5 K.} \textbf{a} Rabi oscillation of spin defects in $\upbeta$-GeS$_2$ by applying the pulse sequence shown in the top panel. \textbf{b} Measurement of the spin coherence time $T_2$ of spin defects in $\upbeta$-GeS$_2$ using a Hahn-echo sequence (top panel). The fitted $T_2$ is 10.03$\pm$1.52 $\upmu$s. \textbf{c} Spin-lattice relaxation time ($T_1$) measurement of spin defects in $\upbeta$-GeS$_2$, and the fitted $T_1$ is 1.20$\pm$0.40 ms. The pulse sequence is shown in the top panel. All measurements were performed under a 27.5-mT magnetic field at 5 K.}
\end{figure*}

Next, we apply pulse ODMR technologies to manipulate these spin defects and examine the coherence time, but failed at room temperature with this sample. 
Considering that the decrease of temperature can efficiently reduce the impact of some decoherence sources such as electron-phonon coupling, we subsequently successfully realize the coherent control of spin defects in $\upbeta$-GeS$_2$ at 5 K and under a 27.5-mT magnetic field (Fig. 3). 
Rabi-oscillation experiments are first performed by using the pulse sequence shown in Fig. 3\textbf{a}. 
Begin with the spin state polarization by a laser pulse, the spin is then manipulated by a resonant microwave pulse with varying lengths, and finally read out by another laser pulse. 
Fig. 3\textbf{a} shows a typical Rabi oscillation fitted by the Equation $A + Be^{-\tau/T_2^*}cos(2\pi f\tau +\phi)$, which provides the duration of $\pi$ and $\pi/2$ microwave pulses employed in the followed pulsed ODMR experiments. 
The coherence time $T_2$ is measured by the Hahn-echo sequence and is derived as 10.03$\pm$1.52 $\upmu$s by an exponential-decay fitting (Fig. 3\textbf{b}). 
This value is two orders of magnitude longer than $\rm V_B^-$ in hBN, and is also the longest coherence time reported for spin defects in 2D materials to date. 
Fig. 3\textbf{c} displays the pulse sequence and result of spin-lattice relaxation time $T_1$. 
By fitting data with an exponential decay, we obtain $T_1$ = 1.20$\pm$0.40 ms. This short $T_1$ indicates that the crystal lattice of this sample is not as good as required due to the immature grow technology of $\upbeta$-GeS$_2$ sample.

In spite of this, we anneal these $\upbeta$-GeS$_2$ samples again in air at 500 $^\circ$C for 24 hours, then using this sample we achieve coherent control at room temperature, and the measured $T_2$ reached 20 $\upmu$s (Fig. 4\textbf{a}). 
Fig. 4\textbf{b} exhibits the pulse sequence and result of room-temperature $T_1$. 
It should be noted that this $T_1$ sequence without microwave (called ``OFF'' mode) \cite{Durand2023Optically,Tetienne2013Spin,Crook2020Purcell,Crook2021Photonic} is different from that at 5 K (called ``ON-OFF'' mode), and we adopt it here because the ``OFF'' mode can accumulate data faster when the longer $T_1$ causes the data-accumulation time much longer. 
Although the measurement methods are different, the extended $T_1$ (tens of milliseconds) at room temperature can clearly show that long time annealing could be efficient way to relax the local structure around defects and reduce the influence from phonons.

\textbf{Theoretical simulation.} Here we carried out DFT calculations to unveil the possible structure of these spin defects. Based on the result of electron probe X-ray micro-analyzer (EPMA) experiment (details are shown in Fig. 5\textbf{a} and will be discussed later), we just consider the native defects, external impurities including oxygen, carbon and nitrogen. The valence electrons of
Ge ($s2p2$) and S ($s2p4$) manifest that native defects can
only have unpaired spin ($S$ = 1/2) with charged state. Among them, only the negatively charge $\rm Ge_S^-$ could produce zero-phonon line (ZPL) at 1.71 eV (725 nm) which is close to our measured large packets. The $\rm C_S^-$ ($S$ = 1/2) and $\rm N_S^0$ ($S$ = 1/2) have ZPL at 2.17 eV and 2.26 eV, respectively, and they might be related to the first packet at 615 nm. The simulated phonon side band (PSB) of $\rm Ge_S^-$ and $\rm C_S^-$ are also shown and the corresponding Huang-Rhys factors are 4.2 and 5.9. The wide PSB makes the ZPL not easy to detect. This might be due to the large electron phonon coupling, and for this optical cavity can be used to enhance the brightness and quantum efficiency.

\textbf{Discussions.}
All the fluorescence spectra of the $\upbeta$-GeS$_2$ samples excited at 532 nm are the same as  shown in Fig. 1\textbf{e} (only the relative proportions of the two packets are slightly different, see Fig. S2 in Supplementary Information), regardless of whether the spin signal can be detected.
This suggests that there are various kinds of color centers in $\upbeta$-GeS$_2$ and the possible structures are complex.
The theoretical prediction of the coherence time $T_2$ of spin defects in $\upbeta$-GeS$_2$ can be up to 4.3 ms \cite{Kanai2022Generalized}, which is much longer than our measurement even after long-time annealing.
We speculate that the main reason is the unwanted spin noise around due to the imperfection of samples. 
Therefore, we further perform the EPMA experiment (as mentioned previously). The result shows that the atom-number ratio of Ge:S in the sample is approximately 1:2 (slightly more S), again indicating the sample is not GeS; and in addition, the sample contains a small amount of oxygen (O), carbon (C) and nitrogen (N), and trace amounts of ferromagnetic iron (Fe), cobalt (Co), and gadolinium (Gd) (see Fig. 5\textbf{a}). 
Hence, the spin defects here still could be decoherenced by the nuclear spin bath from the impurities present in the sample. Besides, the $\upbeta$-GeS$_2$ samples have more than two types of defects containing electron spins, which couple and decoherence with each other \cite{Haykal2022Decoherence}.

The generalized criteria for the preferable properties of materials hosting defect spin qubits consolidated by Weber et al. \cite{Weber2010Quantum} include nuclear-spin-free lattice and availability of high-quality single crystals. 
Although high-quality and high-purity single crystal 2D hBN can be purchased commercially, the dense nuclear spins will be a longstanding issue. 
On the contrary, $\upbeta$-GeS$_2$ has ideal nuclear-spin-free lattice, however commercially available $\upbeta$-GeS$_2$ has notable impurities, which hinders current experimental research. 
We believe with improvement of the synthesis of high-quality, high-purity crystals of $\upbeta$-GeS$_2$, the isolated single-spin color center inside can fully unleash the potential of $\upbeta$-GeS$_2$ in quantum applications. 

\begin{figure}[tp]
\centering
\includegraphics[width=0.5\textwidth]{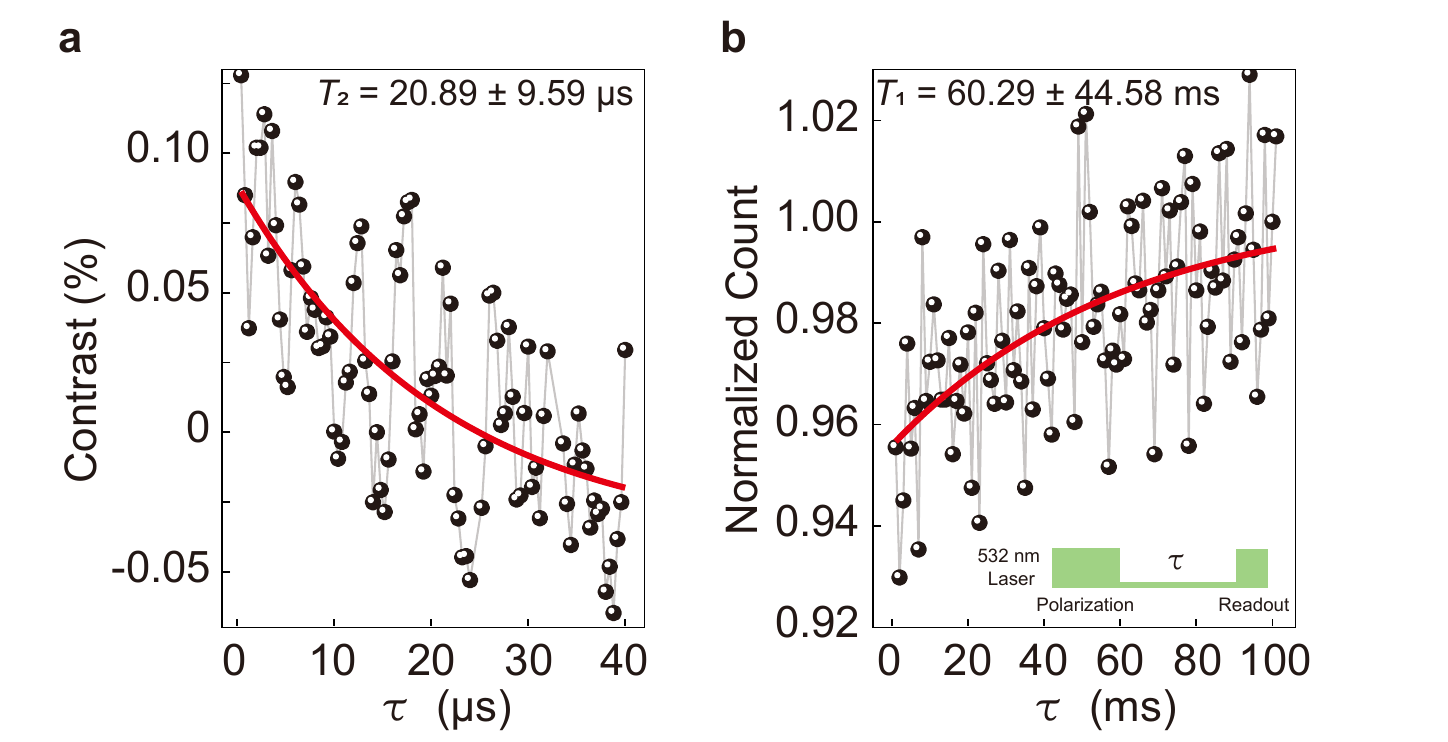}
\caption{\label{Figure 4} \textbf{Room-temperature coherence time (24-hour annealing).} \textbf{a} Room-temperature coherence time $T_2$ = 20.89$\pm$9.59 $\upmu$s and \textbf{b} spin-lattice relaxation time $T_1$ = 60.29$\pm$44.58 ms (the inset is pulse sequence).}
\end{figure}

\begin{figure*}[tp]
\centering
\includegraphics[width=0.9\textwidth]{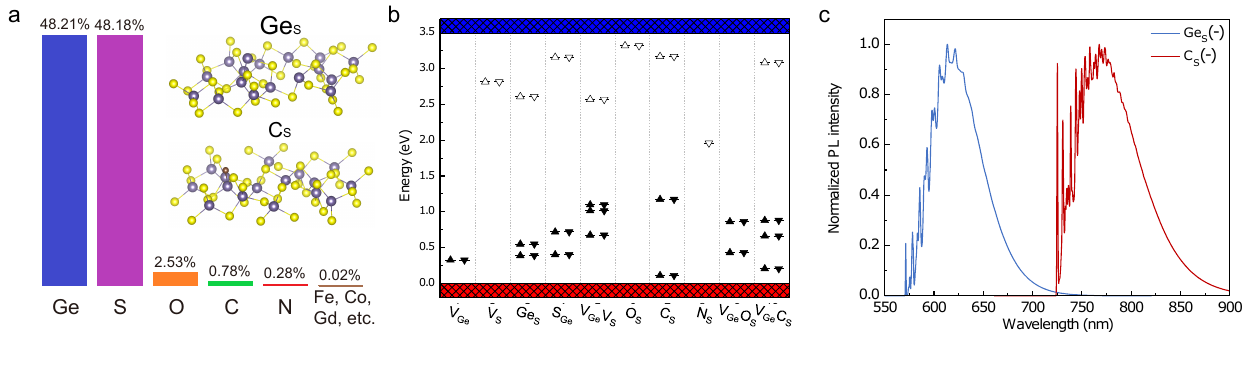}
\caption{\label{Figure 5} \textbf{Defects analysis of the $\upbeta$-GeS$_2$ sample.} \textbf{a} The elemental mass ratio of the $\upbeta$-GeS$_2$ sample was measured by electron probe X-ray micro-analyzer (EPMA). The Ge:S atomic ratio closely matches with the stoichiometry of $\upbeta$-GeS$_2$ rather than GeS. The main impurity atoms are oxygen (O), carbon (C) and nitrogen (N), and the others include some ferromagnetic atoms such as iron (Fe), cobalt (Co) and gadolinium (Gd). \textbf{b} The ground state electronic structure of defects at neutral charge state in $\upbeta$-GeS$_2$. Red and blue indicate the valence and conduction band edge. The triangle indicates spin-up and down channels. \textbf{c} The simulated PSB of Ge$_{\text S}$($-$) and C$_{\text S}$($-$).}
\end{figure*}

\textbf{Conclusion.} To sum up, we report $\upbeta$-GeS$_2$ as another wide-bandgap 2D material hosting optically active spin defects other than hBN.  The nuclear-spin-free lattice allows long coherence time of defect spins, which is unattainable in hBN.
In this work, room-temperature of magnetic-field dependent ODMR spectra and coherent control of the spin defect ensemble in $\upbeta$-GeS$_2$ are demonstrated. With theoretical calculations, we find the substitutional defects could reproduce the measured ZPL with wide PSB. 
The coherence time can reach tens of microseconds (100-folds of that of V$_{\text{B}}^-$ in hBN), the longest among spin defects reported in 2D materials to date. $\upbeta$-GeS$_2$ will be a promising 2D host material for long-coherence-time spin defects.

~\\
\textbf{Methods}
~\\

\textbf{Sample preparation.}
The bulk-crystal $\upbeta$-GeS$_2$ studied in the experiment is commercially available from Shanghai Onway Technology Company, Limited.  
The measured samples are first mechanically exfoliated from bulk crystal by tapes and prepared as the multilayer flakes onto the silicon wafer with 280 nm oxide layer.
Next, these samples on silicon substrates are placed in an annealing furnace and heated uniformly from room temperature to 500 $^\circ$C over 1 hour, then maintained for 1 hour (24 hours) and cooled naturally.
For ODMR measurement, we fabricate 50-$\upmu$m wide and 100-nm thick gold microwave waveguide on the silicon substrates by photolithography and e-beam evaporator \cite{Guo2023Coherent}.
Finally, the annealed $\upbeta$-GeS$_2$ flakes are transferred to these substrates by the dry transfer technology \cite{Kinoshita2019Dry}.
All substrates were cleaned by acetone, isopropyl alcohol and deionized water in sequence before using.

\textbf{Experiment setup.}
Home-built confocal microscopy systems combined with microwave system are used for optical characterization and spin manipulation measurements.
A continuous-wave 532-nm laser modulated by an acousto-optic modulator (AOM) is focused on the samples through a high numerical aperture (N.A.=0.9) objective.
Photoluminescence is collected by the same objective and detected in the emission sideband with a 532-nm long-pass filter by a single photon detector and a data acquisition device.
Scanning is performed by an X-Y-Z nanocube system at room temperature and a X-Y piezo scanning mirror combined with a 4-f system consist of two lenses at low temperature.
The microwave generated by a synthesized signal generator is controlled by a switch, amplified by a microwave amplifier, and finally radiated through the gold waveguide.
An arbitrary signal generator is exploited to produce the corresponding electrical pulse sequences.
A supercontinuum white light picosecond pulsed laser is used to measure the excited-state lifetime.

\textbf{Density functional theory calculations.}
Our calculations are perfomed using the density functional theory (DFT) within the \textit{Vienna ab initio simulation package} (VASP) code~\cite{kresse1996efficiency, kresse1996efficient}. The cutoff energy for the plane wave basis set is 550~eV. The  projector augmented wave (PAW) potentials us used to separate the valence electrons and the core part~\cite{blochl1994projector, kresse1999ultrasoft}. The hybrid density functional of Heyd, Scuseria, and Ernzerhof (HSE)~\cite{heyd2003hybrid} is used to optimize the geometry and calculate electronic structures. The optimized primitive cell is 6.72\AA\ $\times$ 16.08\AA\ $\times$ 11.55\AA\ with 3.5 eV band gap and a $2\times1\times1$ two-layered supercell model is used to avoid the interactions between periodic images and this is sufficient to use the $\Gamma$-point sampling scheme. The interlayer vdW interaction was described with DFT-D3 method of Grimme~\cite{grimme2010consistent}. The atoms are fully relaxed until the forces are less than 0.01~eV/\AA. To calculate the excited state, the $\Delta$SCF method~\cite{gali2009theory} is used. The PSB simulation is done based on the ground and excited state geometries with the overlap between the phonon modes during the excitation~\cite{alkauskas2014first}.

\textbf{Acknowledgements} 

This work is supported by the Innovation Program for Quantum Science and Technology (No. 2021ZD0301200), the National Natural Science Foundation of China (No. 12174370, 12174376, and 11821404, 12304546), the Youth Innovation Promotion Association of Chinese Academy of Sciences (No. 2017492), Anhui Provincial Natural Science Foundation (No. 2308085QA28), China Postdoctoral Science Foundation (No. 2023M733412). This work was partially carried out at the USTC Center for Micro and Nanoscale Research and Fabrication. This work was also supported by the Hungarian National Research, Development and Innovation Office (NKFIH) for Quantum Information National Laboratory of Hungary (grant no.\ 2022-2.1.1-NL-2022-00004), the Horizon Europe EIC Pathfinder QuMicro project (grant no.\ 101046911)
and SPINUS project (grant no.\ 101135699).

\end{document}